\documentstyle[floats,prd,aps,epsfig,eqsecnum,12pt]{revtex}
\makeatletter
\newbox\tempboxa
\newdimen\captionboxsubcount 
\def\capsize#1{\captionboxsubcount=#1pt}
\newdimen\captionboxsub
\captionboxsub=\hsize \advance\captionboxsub by -\captionboxsubcount
\advance\captionboxsub by -\captionboxsubcount
\long\def\@makecaption#1#2{
 \setbox\@tempboxa\hbox{#1 #2}
 \ifdim \wd\@tempboxa >\captionboxsub 
\rightskip=\captionboxsubcount \leftskip=\captionboxsubcount #1 #2 
\else \hbox to\hsize{\hfil\box\@tempboxa\hfil} 
 \fi}
\makeatother
\capsize{30}

\begin{document}
\bibliographystyle{unsrt}
\begin{titlepage}
\begin{flushright}
\begin{minipage}{5cm}
\begin{flushleft}
\small
\baselineskip = 13pt
YCTP-21-98\\
SU-4240-683\\
hep-ph/9808415 \\
\end{flushleft}
\end{minipage}
\end{flushright}
\begin{center}
\Large\bf
Putative Light Scalar Nonet
\end{center}
\vfil
\footnotesep = 12pt
\begin{center}
\large
Deirdre {\sc Black}$^{\it \bf a}$ 
\footnote{Electronic address: {\tt black@physics.syr.edu}}
\quad\quad Amir H. {\sc Fariborz}$^{\it \bf a}$\footnote{Electronic  
address: {\tt amir@suhep.phy.syr.edu}}\\
\vskip 0.5cm
Francesco {\sc Sannino}$^{\it \bf b}$\footnote{
Electronic address : {\tt francesco.sannino@yale.edu}} 
\quad\quad
Joseph {\sc Schechter}$^{\it \bf a}$\footnote{
Electronic address : {\tt schechte@suhep.phy.syr.edu}}\\
{\it 
\qquad $^{\it \bf a}$ Department of Physics, Syracuse University, 
Syracuse, NY 13244-1130, USA.} \\
\vskip 0.5cm
{ \it  \qquad $^{\it \bf b}$ 
Department of Physics, Yale University, New Haven, CT 06520-8120,  
USA.}\\
\end{center}
\vfill
\begin{center}
\bf
Abstract
\end{center}

\begin{abstract}
We investigate the ``family'' relationship of a possible scalar nonet
composed of the $a_0(980)$, the $f_0(980)$ and the $\sigma$ and $\kappa$ type
states found in recent treatments of $\pi\pi$ and $\pi K$ scattering.  We
work in the effective Lagrangian framework, starting from terms which yield
``ideal mixing'' according to Okubo's original formulation.  It is noted
that there is another solution corresponding to dual ideal mixing which
agrees with Jaffe's picture of scalars as $qq\bar q \bar q$ states rather
than $q\bar q$ states.  At the Lagrangian level there is no difference in
the formulation of the two cases (other than the numerical values of the
coefficients).  In order to agree with experiment, additional mass and
coupling terms which break ideal mixing are included.  The resulting model
turns out to be closer to dual ideal mixing than to conventional ideal
mixing; the scalar mixing angle is roughly $-17^{\circ}$ in a
convention where dual ideal mixing is $0^\circ$.  
\baselineskip = 17pt
\end{abstract}

\begin{flushleft}
\footnotesize
PACS number(s): 13.75.Lb, 11.15.Pg, 11.80.Et, 12.39.Fe 
\end{flushleft}
\vfill
\end{titlepage}
\setcounter{footnote}{0}

\section{Introduction}

Recently there has been renewed discussion
\cite{Sannino-Schechter}-\cite{Dmitra} about evidence for low energy broad
scalar resonances in the $\pi\pi$ and $\pi K$ scattering channels.  In the
approach \cite{Sannino-Schechter,Harada-Sannino-Schechter,BFSS} on which
the present paper is based, a need was found for a $\pi\pi$ resonance
($\sigma$) at 560 MeV and a $\pi K$ resonance ($\kappa$) around 900 MeV.
That approach, motivated by the ${1}/{N_c}$ \cite{1n} approximation to
QCD, involves suitably regularized (near the poles) tree level diagrams
computed from a chiral Lagrangian and containing resonances within the
energy range of interest.  Attention is focussed on the real parts which
satisfy crossing symmetry but may in general violate the unitarity bounds.
Then the unknown parameters (properties of the scalars) are adjusted to
satisfy the unitarity bounds (i.e. to agree with experiment).  In this way
an approximate amplitude satisfying both crossing symmetry and unitarity is
obtained.  

Similar results for the scalars have been obtained in different models
\cite{Tornqvist}-\cite{Dmitra} although there is not unanimous agreement.
These are, after all, attempts to go beyond the energy region where chiral
perturbation theory \cite{chp} can provide a practical systematic
framework.  

Now if one accepts a light $\sigma$ and $\kappa$ and notes the
existence of the isovector scalar $a_0(980)$ as well as the $f_0(980)$
there are exactly enough candidates to fill up a nonet of scalars, all
lying below 1 GeV.   Presumably these are not the ``conventional'' p-wave
quark-antiquark scalars but something different.  It would then be
necessary (see for example the discussion on page 355 of \cite{PDG}) to
have an additional nonet of ``conventional'' heavier scalars.

Most mesons fit nicely into a pattern where they have quantum numbers of
quark-antiquark ($q\bar q$) bound states with various orbital angular
momenta.  Furthermore, their masses and decays are (roughly) explained
according to a nonet scheme, first proposed by Okubo \cite{Okubo}, known as
``ideal mixing''.  It has been widely recognized that the low-lying scalars
(at least the well observed $a_0(980)$ and $f_0(980)$) do not appear to fit
this usual pattern.  Hence Jaffe \cite{Jaffe} proposed an attractive scheme,
in the context of the MIT bag model \cite{Chodos}, in which the light
scalars are taken to have a $qq\bar q \bar q$ quark structure (and zero
relative orbital angular momenta).  Other models explaining light scalars
as ``meson-meson'' molecules \cite{Isgur} or as due to unitarity corrections
related to strong meson-meson interactions \cite{Tornqvist,vanBeveren} also
involve four quarks at the microscopic level and may possibly be related.

Our concern in the present paper is to study the nonet structure of the
light scalars based on the approach of
\cite{Sannino-Schechter}-\cite{BFSS}.  There, an effective chiral
Lagrangian treatment was used.  In such a treatment, only the $SU(3)$
flavor properties of the scalars are relevant \cite{Callan}.  At this
level, one would not expect any difference in the formulation of our model
since both Okubo's model and Jaffe's model use nonets with the same $SU(3)$
flavor transformation properties.  In fact, we shall show (in Section II)
that the effective Lagrangian defining ideal mixing in Okubo's scheme has
two ``solutions''.  The one he choses explains the light vector mesons with
a natural quark-antiquark structure.  The other solution is identical to
Jaffe's model of the scalars.  We note that it may be formally regarded as
having a dual-quark dual-antiquark structure, where the dual quark is
actually an anti-diquark. 

The initial appearance is that the four masses of the light nonet
candidates obey the ordering relation [Eq. (\ref{new-nonet-hierarchy})
below] of the dual ideal mixing picture but not the more stringent
requirement of this picture Eq. (\ref{sum-rule-b}).  Furthermore the decay
$f_0(980) \rightarrow \pi\pi$ is experimentally observed but is predicted
to vanish according to ideal mixing.  Thus, it is necessary to consider
some corrections to the ideal mixing model.  When such correction terms are
added [to yield a structure like Eq. ({\ref{mixing-mass-Lag}})] the new
model actually displays two different solutions for the particle
eigenstates corresponding to a given scalar mass spectrum (see the
discussion in Section III) so it becomes unclear as to whether the ordinary
or the dual ideal mixing picture is more nearly correct.  In order to
resolve this question the predictions for the
scalar-pseudoscalar-pseudoscalar coupling constants are first computed for
each of these two solutions.  The five coupling constants needed for $\pi
K$ scattering are found to depend on only two parameters - A and B in
Eq. (\ref{interactions}).  Then (see Section IV) the $\pi K$ scattering is
recalculated taking these two parameters as quantities to be fit.  However
it turns out that both solutions yield equally probable fits to the $\pi K$
scattering amplitudes.  Finally, the question is resolved by noting that
only one of the two solution sets gives results which could be compatible
with the previous \cite{Harada-Sannino-Schechter} $\pi\pi$ scattering
analysis and with the $f_0(980) \rightarrow \pi\pi$ decay rate.

The favored solution is characterized by a scalar $\sigma-f_0$ mixing angle
which is closer to the dual form of ideal mixing than to the usual form.
Using a convention [see Eq.(\ref{mixing-convention})] where an angle
$\theta_s =0$ means dual ideal mixing and $\left| \theta_s \right| =
\frac{\pi}{2}$ means conventional ideal mixing, the favored solution has
$\theta_s \approx -17^\circ$.  It should be noted that this result is based
on an analysis of scalar coupling constants which are related to each other
``kinematically'' but which are related to experiment through ``dynamical''
models of $\pi K$ and $\pi\pi$ scattering.  

Some technical details are put in three Appendixes.  Appendix A contains a
brief discussion of some key features of the $qq\bar q \bar q$ scalars as
expected in the quark model.  Appendix B shows how the needed terms of the
Lagrangian including the scalar nonet may be presented in chiral covariant
form.  Finally Appendix C contains a list of the various
scalar-pseudoscalar-pseudoscalar coupling constants and their relations to
the parameters of our Lagrangian and to the scalar and pseudoscalar mixing
angles.

\section{Scalar Nonet Masses}

For orientation, it may be useful to start off by paraphrasing Okubo's
classic discussion \cite{Okubo} of the ``ideal mixing'' of a meson nonet
field, which we denote as the $3 \times 3$ matrix $N_a^{b}(x)$.  In our
case the field will have $J^{P}=0^+$ rather than $J^{P}=1^-$ as in the
original case.  The notation is such that a lower index transforms under
flavor $SU(3)$ in the same way as a quark while an upper index transforms
in the same way as an antiquark.  In this discussion it is not strictly
necessary to mention the quark substructure of N - only its flavor
transformation property will be of relevance.  This lack of specificity
turns out to be an advantage for our present purpose.  

The ``ideal mixing'' model may be defined by the following mass terms of an
effective Lagrangian density:
\begin{equation}
{\cal L}_{mass} = -a {\rm {Tr}}(NN) - b {\rm {Tr}}(NN{\cal M}),
\label{mass-Lag}
\end{equation}
where a and b are real constants while ${\cal M}$ is the ``spurion matrix''
${\cal M}={\rm diag}(1,1,x)$ , $x$ being the ratio of strange to non-strange quark
masses in the usual interpretation.  Iso-spin invariance is being assumed.
The names of the scalar particles with non-trivial quantum numbers are:
\begin{equation}
N = \left[ \begin{array}{c c c}
N_1^1&a_0^+&\kappa ^+\\
a_0^-&N_2^2&\kappa ^0\\
\kappa^-&{\bar \kappa}^0&N_3^3
\end{array} \right],
\end{equation}
with $a_0^0 = (N_1^1 - N_2^2)/ \sqrt 2$.  There are
two iso-singlet states: the combination $(N_1^1 + N_2^2
+ N_3^3)/ \sqrt 3$ is an $SU(3)$ singlet while $(N_1^1
+ N_2^2 - 2N_3^3)/\sqrt 6$ belongs to an $SU(3)$ octet.  These will in
general mix with each other when $SU(3)$ is broken.  Diagonalizing the
fields in Eq. (\ref{mass-Lag}) yields the diagonal (ideally mixed) states
$(N_1^1 + N_2^2)/ \sqrt 2$ and $N_3^3$.

Now it is easy to read off the particle masses from Eq. (\ref{mass-Lag}) in
terms of $a$, $b$ and $x$.  This information is conveniently described by
the two sum rules:
\begin{equation}
m^{2}\left( a_0 \right) = m^{2}\left( \frac {N_1^1 + N_2^2}{\sqrt 2}\right),
\label{sum-rule-a}
\end{equation}
\begin {equation}
m^{2}\left( a_0 \right) - m^{2}\left( \kappa \right) = m^{2}\left( \kappa
\right) - m^{2}\left( N_3^3 \right).
\label{sum-rule-b}
\end{equation}
There are two characteristically different kinds of solutions, depending on
whether both sides of Eq. (\ref{sum-rule-b}) are positive or negative.
Okubo's original scheme amounts to the choice that both sides of
Eq. (\ref{sum-rule-b}) are negative.  Then

\begin{equation}
m^{2}\left( N_3^3 \right)  > m^{2}\left( \kappa \right) > m^{2}\left( a_0
\right) = m^{2}\left( \frac{N_1^1 + N_2^2}{\sqrt 2} \right).
\label{nonet-hierarchy}
\end{equation}
This is consistent with a quark model interpretation of the composite nonet
field:
\begin{equation}
N_a^b \sim q_a {\bar q}^b,
\label{conventional}
\end{equation}
identifying $q_1, q_2, q_3 = u,d,s$.  Specifically,
Eq. (\ref{conventional}) states that $N_3^3$ is composed of one strange
quark and one strange antiquark, $\kappa$ of one non-strange quark and one
strange antiquark while $a_0$ and $(N_1^1 +
N_2^2)/{\sqrt 2}$ have zero strange content.  Thus the ordering in
Eq. (\ref{nonet-hierarchy}) naturally follows if the strange quark is
heavier than the non-strange quark, as has been well established.  This
ideal mixing picture works well for the vector mesons (with the
reidentifications $N_3^{3} \rightarrow \phi$, $(N_1^1 +
N_2^2)/{\sqrt 2} \rightarrow \omega$, $\kappa \rightarrow K^*$ and $a_0
\rightarrow \rho$) and reasonably well for most of the other observed meson
multiplets (see page 98 of \cite{PDG}).  The exceptions are the low-lying
$0^-$ and $0^+$ nonets.  It is generally accepted that the deviation of
the $0^-$ nonet from this picture can be understood from the special
connection of the pseudoscalar flavor singlet with the $U(1)_A$ anomaly of
QCD.  The case of the $0^+$ nonet has been less clear, in part because the
existence of the scalar states needed to fill up a low-lying nonet has been
difficult to establish.

Now a long time ago, Jaffe \cite{Jaffe} suggested that the low-lying
scalars might have a quark substructure of the form $qq\bar q \bar q$
rather than $q\bar q$.  This model can be put in the identical form as our
previous discussion of Eqs. (\ref{mass-Lag}) - (\ref{sum-rule-b}) by
introducing the ``dual'' flavor quarks (actually diquarks):
\begin{equation}
T_a = \epsilon _{abc} {\bar q}^b {\bar q}^c , \quad \quad {\bar T}^a =
\epsilon^{abc}q_{b}q_{c},
\label{dual-quarks}
\end{equation}
wherein it should be noted that the quark fields are anticommuting
quantities.  Then we should write the scalar nonet as 
\begin{equation}
N_a^b \sim {T_a}{{\bar T}^b} \sim 
\left[ \begin{array}{c c c}
\bar {s} \bar {d} ds &\bar {s} \bar {d} us&\bar {s} \bar {d} ud\\
\bar {s} \bar {u} ds&\bar {s} \bar {u} us&\bar {s} \bar {u} ud\\
\bar {u} \bar {d} ds&\bar {u} \bar {d} us& \bar {u} \bar {d} ud
\end{array} \right].
\label{multiquark-nonet}
\end{equation}

In the present $qq\bar q \bar q$ case both sides of Eq. (\ref{sum-rule-b})
should be taken to be positive.  The
tentative identifications $f_0(980) = (N_1^1 + N_2^2)/{\sqrt 2}$ and
$\sigma = N_3^3$ would then lead to an ordering opposite to that of
Eq. (\ref{nonet-hierarchy}),
\begin{equation}
m^{2}\left( f_0 \right)  = m^{2}\left( a_0 \right) > m^{2}\left( \kappa
\right) >  m^{2}\left( \sigma \right).
\label{new-nonet-hierarchy}
\end{equation}
This is in evident good agreement with the experimentally observed equality
of the $f_0(980)$ and $a_0(980)$ masses.  Furthermore it is seen that the
ordering in Eq. (\ref{new-nonet-hierarchy}) agrees with the number of
underlying (true) strange objects present in each meson according to the alternative
ansatz (\ref{multiquark-nonet}).  

If additional terms {\footnote{We are neglecting a possible term $-e{\rm
Tr}(N{\cal M}){\rm Tr}(N{\cal M})$ which is second order in symmetry breaking.}} are added to the ideal mixing model in Eq. (\ref{mass-Lag}) to yield
\begin{equation}
{\cal L}_{mass} = -a {\rm Tr}(NN) - b {\rm Tr}(NN{\cal M}) - c {\rm
Tr}(N)Tr(N) - d {\rm Tr}(N) {\rm Tr}(N{\cal M}),
\label{mixing-mass-Lag}
\end{equation}
the states $(N_1^1 + N_2^2)/{\sqrt 2}$ and $N_3^3$ will
no longer be diagonal.  The physical states will be some linear combination
of these.  This ``non-ideally mixed'' situation will be seen to be required
in order to explain the experimental pattern of scalar decay modes.  We
would like to stress that, in the effective Lagrangian approach, no more
than the assumption of mass terms like (\ref{mixing-mass-Lag}) is required;
it is not necessary to assume a particular quark substructure for $N_a^b$.
That field may represent a structure like (\ref{conventional}), one like
(\ref{multiquark-nonet}), a linear combination of these or something more
complicated.  Of course, it is still interesting to ask whether the
resulting predictions are closer to those resulting from
(\ref{multiquark-nonet}) or from (\ref{conventional}).

A natural question concerns the plausibility of the ``dual'' ansatz in
Eq. (\ref{multiquark-nonet}), which at first sight seems merely contrived
to yield the ordering in Eq. (\ref{new-nonet-hierarchy}).  Jaffe
\cite{Jaffe} showed that there is a dynamical basis for such
an ansatz in the MIT bag model \cite{Chodos}.  It essentially arises from
the strong binding energy in such a configuration due to a hyperfine
interaction Hamiltonian of the form
\begin{equation}
H_{hf} =  - \Delta {\sum}_{i<j} {\bf S}_i
\cdot {\bf S}_j  {\bf F}_i\cdot {\bf {F}}_j
\label{hyperfine-Ham}
\end{equation}
where $\Delta$ is a positive quantity depending on the quark or antiquark
wave functions.  $\displaystyle{{\bf S} = \frac {\mbox{\boldmath
$\sigma$}}{2}}$ is the spin operator and $\displaystyle{{\bf F} = \frac
{\mbox{\boldmath $\lambda$}}{2}}$ ($\mbox {\boldmath{$\lambda$}}$ are the Gell-Mann
matrices) is the color-spin operator.  The sum is to be taken over each
pair $\left( i,j \right)$ of objects (i.e. $qq$, $\bar q \bar q$ or $q\bar
q$) in the hadron of interest.  Eq. (\ref{hyperfine-Ham}) represents an
approximation to the hyperfine interaction obtained from one gluon exchange
in QCD; it is widely used in both quark model \cite{Close} as well as bag
model treatments of hadron spectroscopy.

Standard application of (\ref{hyperfine-Ham}) to the $\rho - \pi$ and
$\Delta -N$ mass differences in the simple quark model yields:
\begin{equation}
\begin{array}{c c}
\langle \pi|H_{hf}|\pi \rangle = -\Delta _{q\bar q}, \quad & \quad \langle
\rho|H_{hf}|\rho \rangle =
+\frac{1}{3}\Delta _{q\bar q},\\
\langle N|H_{hf}|N \rangle = - \frac{1}{2}\Delta _{qqq},\quad  & \quad
\langle \Delta|H_{hf}|\Delta \rangle =
+\frac{1}{2}\Delta _{qqq},
\end{array}
\label{hf-mass-splitting}
\end{equation}
in which a subscript has been given to the $\Delta$ factor for each quark
configuration.  It can be seen that $\Delta$ is expected to be fairly
substantial - of order of several hundred MeV - in these cases.  The
evaluation of the expectation value of Eq. (\ref{hyperfine-Ham}) for the
lowest scalar $qq\bar q \bar q$ nonet state \cite{Jaffe} is more
complicated than for the above cases and yields a large enhancement factor
due to the color and spin Clebsch-Gordon manipulations:
\begin{equation}
\langle 0^+|H_{hf}|0^+ \rangle \approx -2.71 \Delta_{qq\bar q \bar q}.
\label{exotic-hf-mass-splitting}
\end{equation}

Thus, quark model arguments make plausible a strongly bound $qq\bar q \bar
q$ configuration.  It should be remarked that the lowest lying $0^+$ nonet
state in the quark model which diagonalizes Eq. (\ref{hyperfine-Ham}) is a
particular linear combination of state 1 in which the $qq$ pair is in a
$\bar 3$ of color and is a spin singlet and state 2 in which the $qq$ pair
is in a color $6$ and is a spin triplet:
\begin{equation}
|0^+ \rangle \approx 0.585 |1 \rangle + 0.811 |2 \rangle.
\label{exotic-mass-eigenstate}
\end{equation}

A derivation of Eqs. (\ref{exotic-hf-mass-splitting}) and
(\ref{exotic-mass-eigenstate}) is given in Appendix A.

\section{Scalar nonet mixings and trilinear couplings}

First let us consider the consequences of the generalized mass terms
(\ref{mixing-mass-Lag}), which allow for arbitrary deviations from ideal
mixing.  The squared masses of the $a_0$ and $\kappa$ are read off as 
\begin{eqnarray}
m^2 \left( a_0 \right) &=& 2a +2b \nonumber \\
m^2 \left( \kappa \right) & =& 2a + \left( 1+x \right) b.
\label{masses}
\end{eqnarray}

Using the basis $\left( N_3^{3}, \quad \frac {N_1^1 + N_2^2}{\sqrt 2}
\right)$, the mass squared matrix of the two iso-scalar mesons is also read
off as 
\begin{equation}
\left[ \begin{array}{c c}
2m^2 \left( \kappa \right) - m^2 \left( a_0 \right) +2c+2dx \: & \sqrt{2} \left[ 2c + \left( 1+x \right)d \right]\\ \\ 
\sqrt{2} \left[ 2c + \left( 1+x \right)d \right] \: & m^2 \left( a_0 \right) +4c+4d
\end{array} \right].
\label{mass-matrix}
\end{equation}
In obtaining this result Eqs. (\ref{masses}) were used to eliminate the
parameters $a$ and $b$.  The physical isoscalar states and squared masses
are to be obtained by diagonalizing this matrix.  Notice that the four
parameters $a$, $b$, $c$ and $d$ may be essentially traded for the four
masses.  We will take \cite{Harada-Schechter} the strange to non-strange
quark mass ratio $x$ to be $20.5$ for definiteness.  Then, up to a discrete
ambiguity, the mixing angle between the two isoscalars will be predicted.

It seems worthwhile to point out that the structure of our mass formulas
provides {\it {constraints}} on the allowed masses.  To see this, note
that the diagonalization of (\ref{mass-matrix}) yields the following
quadratic equation for $\tilde {d} = \left( 1-x \right) d$:
\begin{eqnarray}
6 {\tilde {d}}^2 &-& 8 \left[ m^2\left( a_0 \right) - m^2 \left( \kappa
\right) \right] {\tilde d}  \nonumber \\
&+& \left[ 3m^2 \left( \sigma \right) m^2 \left(
f_0 \right) - 6 m^2 \left( \kappa \right) m^2 \left( a_0 \right) + 3m^4
\left( a_0 \right)  - \delta \left( 4m^2\left( \kappa \right) - m^2 \left(
a_0 \right) \right ) \right] = 0,
\label{dtilde}
\end{eqnarray}
where $\delta = m^2 \left( \sigma \right) + m^2 \left( f_0 \right) - 2m^2
\left( \kappa \right) $ and we have eliminated $c$ according to $6c = \delta
- \left( 4 + 2x \right) d$.  Here $\sigma$ and $f_0$ stand respectively
for the lighter and heavier isoscalar particles.  In order for $\tilde d$
to be purely real, required at the present level of analysis, we must have 
\begin{eqnarray}
\left[ m^2 \left( a_0 \right) - 4 m^2 \left( \kappa \right) \right] ^2 +
3 m^2 \left( a_0 \right) \left[ m^2 \left( \sigma \right) + m^2 \left( f_0
\right) \right] &+& 9 m^2 \left( \sigma \right) m^2 \left( f_0 \right) 
\nonumber \\ &<& {12 m^2 \left( a_0 \right) \left[ m^2 \left( \sigma \right) +
m^2 \left( f_0 \right) \right]}.
\label{reality-condition}
\end{eqnarray}
Taking $m\left( f_0 \right) = 980$MeV and $m\left( a_0 \right) = 983.5$ MeV,
according to \cite{PDG}, and $m\left( \sigma \right) = 550$ MeV from
\cite{Harada-Sannino-Schechter} we find that (\ref{reality-condition})
limits the allowed range of $m\left( \kappa \right)$ to
\begin{equation}
685 \: {\rm MeV} < m\left( \kappa \right) < {980 \: {\rm MeV}}.
\label{kappa-bound}
\end{equation}
It is encouraging that our recent study of $\pi K$ scattering \cite{BFSS}
(see also \cite{Ishida_kappa}) yielded a value for $m\left( \kappa \right)$ of
about $900$MeV, within this range.

The physical particles $\sigma$ and $f_0$ which diagonalize
(\ref{mass-matrix}) are related to the basis states $N_3^3$ and $(N_1^1 +
N_2^2)/{\sqrt 2}$ by
\begin{equation}
\left( \begin{array}{c} \sigma\\ f_0 \end{array} \right) = \left(
\begin{array}{c c} {\rm cos} \theta_s & -{\rm sin} \theta_s \\ {\rm sin}
\theta_s & {\rm cos} \theta_s \end{array} \right) \left( \begin{array}{c}
N_3^3 \\ \frac {N_1^1 + N_2^2}{\sqrt 2} \end{array} \right),
\label{mixing-convention}
\end{equation}
which defines the scalar mixing angle $\theta_s$.  Since Eq. (\ref{dtilde})
for $\tilde d$ is quadratic we expect two different solutions for the pair
$\left( c, \: d \right)$ and hence for $\theta_s$ when we fix the four
scalar masses $m\left( a_0 \right)$, $m\left( \kappa \right)$, $m\left(
\sigma \right)$ and $m\left( f_0 \right)$.  A numerical diagonalization for
the choice $m\left( \kappa \right) \approx 900$ MeV as above yields the two
possible solutions
\begin{eqnarray}
\left( a \right)  \quad &{\theta}_s& \: \approx \: -21^{\circ} \\ \nonumber
\left( b \right) \quad &{\theta}_s& \: \approx \: -89^{\circ}.
\label{mixing-angles}
\end{eqnarray}
Solution ($a$) corresponds to a $\sigma$ particle which is mostly $N_3^3$
(presumably $qq\bar q \bar q$ type) while solution ($b$) corresponds to
$\sigma$ which is $(N_1^1 + N_2^2)/{\sqrt 2}$ (i.e. $q\bar q$ type).
We see that when deviations from ideal mixing are allowed, the pattern of
low lying scalar masses is by itself not sufficient to determine the quark
substructure of the scalars.  This statement is based on
(\ref{mixing-mass-Lag}) which contains all terms at most linear in the mass
spurion ${\cal M}$.

\begin{figure}
%\vspace{2.5in}
\centering
\epsfig{file=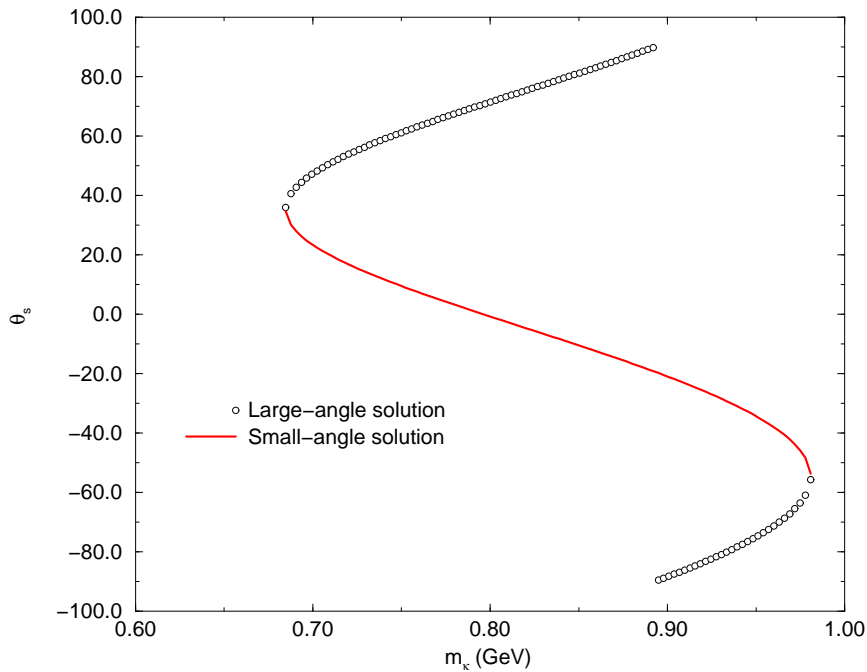, height=5in, angle=270}
\caption
{Scalar mixing angle solutions as functions of $m_\kappa$.}
\label{theta-fig}
\end{figure}

For the complete allowed range of $m_{\kappa}^2$ in Eq. (\ref{kappa-bound})
the two (``small'' and ``large'') mixing angle solutions are displayed in
Fig. \ref{theta-fig}.  Notice that the small angle solution is zero for
$m_{\kappa} \approx 800$ MeV; this is approximately where $c=d=0$, which
would correspond to the dual ideal mixing situation.  In our convention
$-\frac {\pi}{2} \leq {\theta_s} \leq \frac {\pi}{2}$.

Next let us consider the trilinear scalar-pseudoscalar-pseudoscalar
interaction which is related to the main decay modes of the light scalar
nonet states.  We denote the matrix of pseudoscalar nonet fields  by
$\phi_a^b \left( x \right)$.  The general $SU\left( 3 \right)$ flavor
invariant $N\phi \phi$ interaction is written as 
\begin{eqnarray}
{\cal L}_{N\phi \phi} =
A{\epsilon}^{abc}{\epsilon}_{def}N_{a}^{d}{\partial_\mu}{\phi}_{b}^{e}{\partial_\mu}{\phi}_{c}^{f}
&+& B {\rm Tr} \left( N \right) {\rm Tr} \left({\partial_\mu}\phi
{\partial_\mu}\phi \right) + C {\rm Tr} \left( N {\partial_\mu}\phi \right) {\rm
Tr} \left( {\partial_\mu}\phi \right) \nonumber \\
 &+& D {\rm Tr} \left( N \right) {\rm Tr}
\left({\partial_\mu}\phi \right)  {\rm Tr} \left( {\partial_\mu}\phi \right),
\label{interactions}
\end{eqnarray}
where $A, B, C, D$ are four real constants.  The derivatives of the
pseudoscalars were introduced in order that (\ref{interactions}) properly
follows from a chiral invariant Lagrangian in which the field $\phi_a^b$
transforms non-linearly under axial transformations.  The chiral aspect of
our model is largely irrelevant to the discussion in the present paper but,
for completeness, will be briefly treated in Appendix B.  

Notice that the first term of (\ref{interactions}) may be rewritten as 
\begin{eqnarray}
2A {\rm Tr} \left( N{\partial_\mu}{\phi}{\partial_\mu}{\phi}\right) - A
{\rm Tr} \left( N \right) {\rm Tr} \left({\partial_\mu}\phi
{\partial_\mu}\phi\right) 
 &-& 2A {\rm Tr} \left( N {\partial_\mu}\phi \right)
{\rm Tr} \left( {\partial_\mu}\phi \right) \nonumber \\
 &+& A {\rm Tr} \left( N \right)
{\rm Tr} \left({\partial_\mu}\phi \right) {\rm Tr} \left(
{\partial_\mu}\phi \right).
\end{eqnarray}
Thus, if desired, the complicated looking first term of (\ref{interactions})
may be eliminated in favor of the most standard form ${\rm Tr} \left(
N{\partial_\mu}{\phi}{\partial_\mu}{\phi}\right)$.  Our motivation for
presenting it in the way shown is that, by itself, the first term of  
(\ref{interactions}) predicts zero coupling constants for both $f_0
\rightarrow \pi\pi$ and $\sigma \rightarrow K\bar K$ when the ``dual''
ideal mixing identifications, $\sigma = N_3^3$ and $f_0 = (N_1^1 +
N_2^2)/{\sqrt 2}$, are made.  This is in agreement with Jaffe's picture (see
Section VB of \cite{Jaffe}) of the dominant scalar decays arising as the
``falling apart'' or ``quark rearrangement'' of their constituents.  It is
easy to see from (\ref{multiquark-nonet}) that $N_3^3$ cannot fall apart
into $K\bar K$ and that $(N_1^1 + N_2^2)/{\sqrt 2}$ cannot fall apart
into $\pi\pi$.

Of course $f_0\rightarrow \pi\pi$ must be non-zero because $f_0(980)$ is
observed in $\pi\pi$ scattering.  In fact it also vanishes with just
the term ${\rm Tr}\left( N \partial_\mu \phi \partial_\mu \phi \right)$ and
the ``conventional'' identification $\sigma = (N_1^1 + N_2^2)/{\sqrt 2}$
and $f_0 = N_3^3$.  Our model contains two sources for $f_0 \rightarrow \pi
\pi$: the deviation from ideal mixing due to the $c$ and $d$ terms in
(\ref{mixing-mass-Lag}) and also the presence of more than one term in
(\ref{interactions}).  Note again that the use of (\ref{mixing-mass-Lag})
and (\ref{interactions}) does not require us to make any commitment as to
the quark substructure of $N_a^b$. 

Using isotopic spin invariance, the trilinear $N\phi\phi$ interaction
resulting from (\ref{interactions}) must have the form

\begin{eqnarray}
-{\cal L}_{N\phi\phi} &=& \frac{\gamma_{\kappa K \pi}}{\sqrt 2} \left(
\partial_\mu {\bar K} \mbox{\boldmath ${\tau}$} \cdot \partial_\mu {\mbox{\boldmath ${\pi}$}}
\kappa + h.c. \right) + \frac{\gamma_{\sigma \pi \pi}}{\sqrt 2}
\sigma \partial_\mu \mbox{\boldmath ${\pi}$} \cdot \partial_\mu
{\mbox{\boldmath ${\pi}$}} \nonumber \\
&+& \frac{\gamma_{\sigma K  K}}{\sqrt 2}
\sigma \partial_\mu {\bar K} \partial_\mu {K}
+ \frac{\gamma_{f_0 \pi \pi}}{\sqrt 2}
f_0 \partial_\mu \mbox{\boldmath ${\pi}$} \cdot \partial_\mu \mbox{\boldmath ${\pi}$}
+ \frac{\gamma_{f_0 K K}}{\sqrt 2}
f_0 \partial_\mu {\bar K} \partial_\mu {K} \nonumber \\
&+& \frac{\gamma_{a_0 K  K}}{\sqrt 2} \partial_\mu {\bar K} \mbox{\boldmath ${\tau}$} \cdot {\bf a_0} \partial_\mu {K} +
\gamma_{\kappa {K} \eta} \left(
{\bar \kappa} \partial_\mu K \partial_\mu {\eta} + h.c. \right) +
\gamma_{\kappa {K} \eta '} \left(
{\bar \kappa} \partial_\mu K \partial_\mu {\eta '} + h.c. \right)
\nonumber \\
&+& \gamma_{a_0 \pi\eta} {\bf a_0} \cdot \partial_\mu \mbox{\boldmath ${\pi}$} \partial_\mu \eta + 
 \gamma_{a_0 \pi\eta'} {\bf a_0} \cdot \partial_\mu \mbox{\boldmath
${\pi}$} \partial_\mu \eta'  \nonumber \\
&+& \gamma_{\sigma \eta \eta} \sigma \partial_\mu \eta \partial_\mu \eta
+ \gamma_{\sigma \eta \eta'} \sigma \partial_\mu \eta \partial_\mu \eta'
+ \gamma_{\sigma \eta' \eta'} \sigma \partial_\mu \eta' \partial_\mu
\eta' \nonumber \\
&+&  \gamma_{f_0 \eta \eta} f_0 \partial_\mu \eta \partial_\mu \eta
+  \gamma_{f_0 \eta \eta'} f_0 \partial_\mu \eta \partial_\mu \eta'
+ \gamma_{f_0 \eta' \eta'} f_0 \partial_\mu \eta' \partial_\mu \eta',
\label{trilinear-interactions}
\end{eqnarray}
where the $\gamma$'s are the coupling constants.  The fields which appear
in this expression are the isomultiplets:
\begin{eqnarray}
K= \left( \begin{array}{c} K^+ \\ K^0 \end{array} \right),
\quad {\bar K}= \left( \begin{array}{cc} K^- & {\bar K^0} \end{array}
\right)&,& \quad
\kappa = \left( \begin{array}{c} \kappa ^+ \\ \kappa^0 \end{array}
\right), \quad
{\bar \kappa} = \left( \begin{array}{cc} {\kappa}^- & {\bar {\kappa}^0}
\end{array} \right),  \nonumber \\
\pi^\pm = \frac{1}{\sqrt 2} \left( \pi_1 \mp i\pi_2 \right)&,& \quad
\pi^0 = \pi_3,  \nonumber \\
a_0^\pm = \frac{1}{\sqrt 2} \left( a_{01} \mp ia_{02} \right)&,& \quad
a_0^0 = a_{03},
\label{isomultiplets}
\end{eqnarray}
in addition to the isosinglets $\sigma$, $f_0$, $\eta$ and $\eta'$.  The
expressions for the $\gamma$'s in terms of the parameters $A$, $B$, $C$ and
$D$ as well as the scalar and pseudoscalar mixing angles are listed,
together with some related material, in Appendix C.  Notice that if we
restrict attention to those terms in which neither an $\eta$ nor an $\eta'$
appear [first six terms of (\ref{trilinear-interactions})], their coupling
constants only involve two parameters $A$ and $B$.  These are the terms which
will be needed for the subsequent work in the present paper.

\section{Testing the model's coupling constant predictions}

Now let us consider how well the five coupling constants $\gamma_{\kappa
{K} \pi}$, $\gamma_{\sigma \pi \pi}$, $\gamma_{\sigma K K}$, $\gamma_{f_0
\pi \pi}$ and $\gamma_{f_0 K K}$, can be correlated in terms of the two
parameters $A$ and $B$.  These coupling constants, which are listed in
Eqs. (\ref{kappa-K-pi}-\ref{f-K-barK}) are the ones which are relevant for the discussions of $\pi
\pi$ scattering given in \cite{Harada-Sannino-Schechter} and $\pi K$
scattering given in \cite{BFSS}.

A very important question concerns the way in which these $\gamma$'s are to
be related to experiment.  For an ``isolated'' narrow resonance the
magnitude of the coupling constant is proportional to the square root of the
width.  Actually, the only one of the five for which this prescription
roughly applies is $\gamma_{f_0\pi\pi}$; the appropriate formula is given
in Eq. (4.5) of \cite{Harada-Sannino-Schechter}.  Even here there is a
practical ambiguity in that, while the $\pi\pi$ branching ratio is listed
in \cite{PDG}, the total width is uncertain in the range $40-100$MeV.  The
determination $\left| \gamma_{f_0\pi\pi} \right| = 2.43$ ${\rm GeV}^{-1}$
given in \cite{Harada-Sannino-Schechter} is based on using $\Gamma_{tot}
\left( f_0 \right)$ as a parameter in the model analysis of $\pi\pi$
scattering and making a best fit.  

The situation for $\gamma_{f_0 K K}$ is somewhat similar due to the
poorly determined $\Gamma_{tot} \left( f_0 \right)$.  There is an
additional difficulty since the central value of the $f_0(980)$ mass is
{\it {below}} the $K\bar K$ threshold.  Thus the value $\left|
\gamma_{f_0 K K} \right| \approx 10$ ${\rm GeV}^{-1}$ presented in
Section V of \cite{Harada-Sannino-Schechter}, is based on a model taking
the finite width of the initial state into account.  Incidentally, the
non-negligible branching ratio for $f_0 \rightarrow K\bar K$ in spite of
the unfavorable phase space is an indication that the $f_0$
``wavefunction'' has an important piece containing $s\bar s$.  

The $\sigma$, as ``seen'' from the analysis of
\cite{Harada-Sannino-Schechter}, for example, is neither isolated nor
narrow.  A suitable regularization of the tree amplitude near the $\sigma$
pole was argued to be of the form:
\begin{equation}
\frac{m_{\sigma}G}{m_{\sigma}^2 - s} \rightarrow
\frac{m_{\sigma}G}{m_{\sigma}^2 - s - im_\sigma G^{\prime}},
\label{regularization}
\end{equation}
where $G$ and $G^{\prime}$ are real.  G is taken to be proportional to
$\gamma_{\sigma\pi\pi}^2$ while $G^{\prime}$ is considered to be a
regularization parameter.  For a narrow resonance with negligible
background it would be expected that $G^{\prime}=G$.  However, considering
both G and $G^ \prime$ as quantities to be fit (or essentially equivalently,
restoring local unitarity in a crossing symmetric way) yields $G^{\prime}
\neq G$.  The determination $\left| \gamma_{\sigma\pi\pi} \right| =
7.81$ ${\rm GeV}^{-1}$ is based on such a fit.

The situation concerning $\gamma_{\kappa {K} \pi}$ is similar to the
one for $\gamma_{\sigma\pi\pi}$.  Making an analogous fit to the
$\displaystyle{I=\frac{1}{2}}$ amplitude of $\pi K$ scattering (see Section
IV of \cite{BFSS}) yields $\left|
\gamma_{\kappa {K} \pi} \right| \approx 5$ ${\rm GeV}^{-1}$.  This value,
however, is based on inputting the $\left| \gamma_{f_0\pi\pi} \right|$, 
$\left| \gamma_{f_0 K K}\right|$ and $\left| \gamma_{\sigma\pi\pi}
\right|$  values obtained as above and making a particular choice of
$\gamma_{\sigma K K}$.  The value of $\gamma_{\sigma K K}$ was
however not very accurately determined in this model; a compromise choice
was $\gamma_{\sigma K K} \approx 8 \: {\rm GeV}^{-1}$.  

A summary of the coupling constants previously obtained is shown in
Table \ref{couplings-table}.

\begin{table}
\begin{tabular}{lcc}
coupling constant & value (${\rm GeV}^{-1}$) & obtained from \\
\tableline
$\left| \gamma_{f_0 \pi\pi} \right|$ & 2.4 & $\pi\pi$ scattering \\
$\left| \gamma_{f_0 K K}\right|$ & $\approx 10$ & $\pi\pi$ scattering \\
$\left| \gamma_{\sigma \pi\pi} \right|$ & 7.8  & $\pi\pi$ scattering \\
$\left| \gamma_{\kappa K \pi} \right|$ & 5.0 & $\pi K$ scattering \\
$ \gamma_{\sigma K K}$ & $\approx 8$ & $\pi K$ scattering \\
\end{tabular}
\caption{Coupling constants previously obtained in [2] and [3].}
\label{couplings-table}
\end{table}

The discussion above illustrates that it seems necessary to obtain the
coupling constants of the low-lying scalars from a detailed consideration
of the relevant scattering processes.  It is not sufficient to read them
off from \cite{PDG} at the present time.  Furthermore their interpretation
is linked to the dynamical model from which they are obtained.  

It seems to us that a relatively clean way to test the correlation between
the coupling constants in Table \ref{couplings-table} is to recalculate the
$\pi K$ scattering amplitude and, instead of taking $\left| \gamma_{f_0\pi\pi}
\right|$, $\left| \gamma_{f_0 K K}\right|$ and $\left|
\gamma_{\sigma\pi\pi} \right|$ from the $\pi\pi$ scattering output and
regarding $\gamma_{\kappa{K} \pi}$ and $\gamma_{\sigma K K}$ as fitting
parameters as in \cite{BFSS}, just $A$ and $B$ are now taken to be fitted.

We work within the same theoretical framework that was developed in
\cite{Harada-Sannino-Schechter} for the $\pi\pi$ scattering analysis and was
further explored in \cite{BFSS} for the case of $\pi K$ scattering. In this
framework, the $\pi K$ scattering amplitude is computed in a model
motivated by the ${1}/{N_c}$ picture of QCD and its real
part is given as a sum of regularized tree level graphs which include all
resonances that contribute to the amplitude up to the energy region of
interest.  The relevant Feynman diagrams are shown in Fig. 1 of
\cite{BFSS}.

In the $\displaystyle{I=\frac{1}{2}}$ channel, we perform a $\chi^2$ fit,
using the MINUIT package, of this model to the experimental data.
Specifically, in addition to $A$ and $B$, the parameters to be fit are the
regularization parameter in the $\kappa$ propagator, $G^{\prime}_\kappa$
(which can also be interpreted as a total $\kappa$ decay width), and
parameters of the resonance $K_0^*(1430)$: its mass $M_*$, its coupling
$\gamma_*$ and the regularization parameter in its s-channel propagator
$G'_*$.  This will be done for different choices of $m_\kappa$.  Note that
the scalar mixing angle $\theta_s$ (see Section III) will be different for
each choice of $m_\kappa$.  In fact, as already discussed, this actually
gives two different mixing angles for each $m_\kappa$, one (large angle
solution) closer to the $q\bar q$ ansatz (\ref{conventional}) and the other
(small angle solution) closer to the $qq\bar q\bar q$ ansatz
(\ref{multiquark-nonet}).  It is very interesting to see which one is
chosen in our model. More details of the model are given in \cite{BFSS}.
The possible values of $m_\kappa$ are limited by (\ref{kappa-bound}) for
consistency with our present model for masses based on
Eq. (\ref{mixing-mass-Lag}).

\begin{figure}
%\vspace{2.5in}
\centering
\epsfig{file=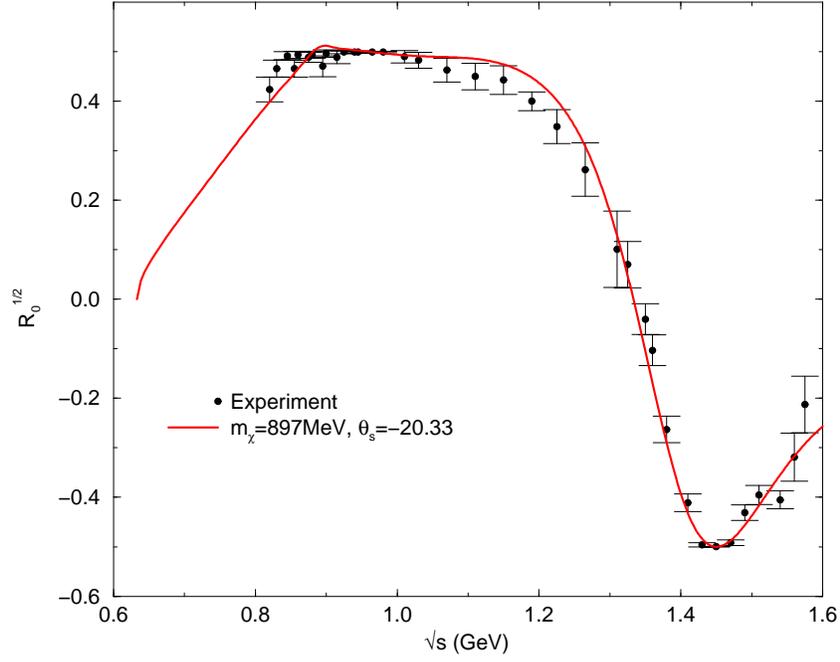, height=5in, angle=270}
\caption
{
Comparison of the theoretical prediction of $R_0^{1/2}$ with its
experimental data.
}
\label{fig-I12}
\end{figure}

Let us first choose $m_\kappa = 897$MeV, as obtained in \cite{BFSS}.  Then
the fit {\footnote{The experimental data points are taken from
\cite{Aston}.}} to the real part of the
$\displaystyle{I=\frac{1}{2}}$ amplitude, ${\rm R}_0^{\frac {1}{2}}$ is
shown in Fig. \ref{fig-I12} while the fitted parameters and resulting
predicted coupling constants are given in Table \ref{Fit_897}.  The results
for both possible mixing angles corresponding to $m_\kappa = 897$MeV are
included.  It is seen that the $\chi^2$ fits to ${\rm R}_0^{\frac {1}{2}}$
are essentially equally good compared to each other and compared to the one
in \cite{BFSS}.  However if we compare the coupling constants in Table
\ref{Fit_897} with those obtained previously in Table \ref{couplings-table}
we see that while the coupling constants $\gamma_{f_0\pi\pi}$, $\gamma_{f_0
K K}$, $\gamma_{\sigma\pi\pi}$ and $\gamma_{\kappa K \pi}$ obtained with
$\theta_s \approx -20^{\circ}$ agree with those obtained earlier in
connection with $\pi\pi$ and $\pi K$ scattering, their values obtained with
$\theta_s \approx -89^{\circ}$ do not agree so well.

Furthermore the value of $\gamma_{f_0\pi\pi}$ obtained with $\theta_s
\approx -89^{\circ}$ would lead to a value for the $f_0$ width several
times larger than the experimentally allowed range.  It thus seems that the
$qq\bar q\bar q$ picture, to which $\theta_s \approx -17^{\circ}$ is much
closer, gives a better overall description of the scalar nonet than does
the $q\bar q$ picture.

It is interesting to investigate the effect of changing $m_\kappa$ within
the range given in Eq. (\ref{kappa-bound}).  As examples, Tables
\ref{Fit_875} and \ref{Fit_800} show the fitted parameters for $m_\kappa =
875$ MeV and $m_\kappa = 800$ MeV respectively.  Several trends can be
discerned.  As $m_\kappa$ decreases from 897 MeV the goodness of fit
actually improves from $\chi^2 = 3.94$ to $\chi^2 = 2.3$ at $m_\kappa =
800$ MeV.  On the other hand the value of $\gamma_{f_0 \pi\pi}$ increases
so that at $m_\kappa = 875$ MeV the $f_0 \rightarrow \pi\pi$ width is in
slightly better agreement with experiment and at $m_\kappa = 800$ MeV it is
many times larger than allowed by experiment.  It seems that the fit at
$m_\kappa = 875$ MeV is not very different from the one at $m_\kappa = 897$
MeV; this gives an estimate of the ``theoretical uncertainty'' in our
calculation.  On the other hand $m_\kappa = 800$ MeV seems to be ruled out,
as are still lower values of $m_\kappa$. 

\begin{figure}
%\vspace{2.5in}
\centering
\epsfig{file=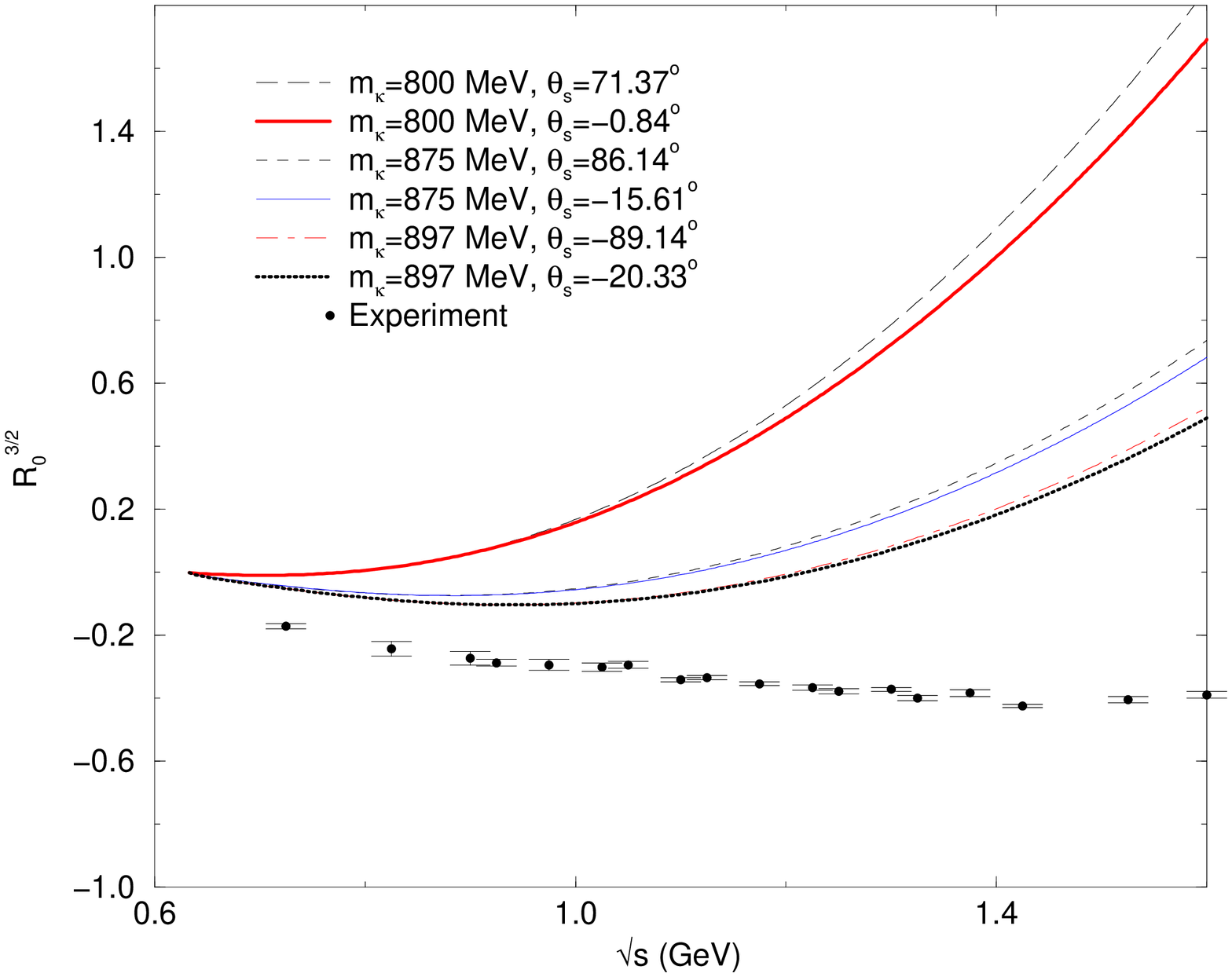, height=5in, angle=270}
\caption
{
Comparison of the theoretical predictions of $R_0^{3/2}$ with its
experimental data.
}
\label{fig-I32}
\end{figure}

Another argument in favor of the larger values of $m_\kappa$ can be made by
examining the $\displaystyle{I=\frac{3}{2}}$ $\pi K$ amplitude
{\footnote{The experimental data points are taken from \cite{Estabrooks}.}},
shown in Fig. \ref{fig-I32}.  It is seen that decreasing $m_\kappa$ worsens
the agreement with experiment.  This feature arises because
$\gamma_{\sigma K K}$, to which the $\displaystyle{I=\frac{3}{2}}$
amplitude is sensitive, increases with decreasing $m_\kappa$.  This
situation was discussed in more detail in section V of \cite{BFSS}, where
it was noted that higher mass resonances may be important in this channel.

We note that the three parameters describing the $K_0^*(1430)$ are stable to
varying $m_\kappa$.

All the fits yield for the parameters $A$ and $B$ that
$\frac {B}{A} {\small >} \!\!\!\! {\small _ \sim} -1$.  Using (\ref{interactions}) then shows
that ${\cal L}_{N\phi\phi}$ approximately looks like 
\begin{equation}
{\cal L}_{N\phi\phi} \approx 2A \left[ {\rm Tr}\left( N \partial_\mu \phi
 \partial_\mu \phi \right) - \rho {\rm Tr}\left( N \right) {\rm Tr} \left(
\partial_\mu \phi \partial_\mu \phi \right) \right] + \cdot \cdot \cdot,
\label{approx-scalar-Lag}
\end{equation}
where $\rho$ is a positive number slightly less than unity and the three
dots stand for the $C$ and $D$ terms which only contribute to vertices
involving at least one $\eta$ or $\eta'$. 

Using this model we can also estimate the partial decay width of $a_0(980)
\rightarrow K{\bar K}$ which is entirely determined in terms of the
parameter $A$ [see Eq.(\ref{a_0-coupling})].  As in the case of $f_0(980)$, the
resonance lies below the decay threshold so the effect of the finite width of
the decaying state must be taken into account [see for example footnote 2
of \cite{Harada-Sannino-Schechter}].  The results are shown in Table V
(taking $m_\kappa = 897$ MeV) corresponding to the extremes of the total
width range given in \cite{PDG}.  Also the effect of the mass difference
between the charged and neutral kaons is taken into account.  The numerical
values seem reasonable.

\begin{table}
\begin{center}
$\begin{array}{|c|c|c|}
\hline
{\rm Fitted \hskip .2cm Parameter}  & {\rm \theta_s=-20.33} & {\rm
\theta_s=-89.14}
\\ \hline 
   G'_\kappa   &   314 \pm 3 \: {\rm MeV}        &   322 \pm 3 \: {\rm MeV}  
\\ \hline
    M_*        &   1390 \pm 4 \: {\rm MeV}       & 1389 \pm 4 \: {\rm MeV}
\\ \hline
   \gamma_*    & 4.42 \pm 0.09\: {\rm GeV}^{-1}   & 4.4 \pm 0.09 \: {\rm GeV}^{-1}
\\ \hline
   G'_*        &  275 \pm 10 \: {\rm MeV}        & 273 \pm 11  \: {\rm MeV}
\\ \hline
   A           & 2.51 \pm 0.03  \: {\rm GeV}^{-1}  & 2.57 \pm 0.03 \: {\rm GeV}^{-1}
\\ \hline
   B           &-1.95 \pm 0.04  \: {\rm GeV}^{-1}  & -2.12 \pm 0.04  \: {\rm GeV}^{-1}
\\ \hline 
  \chi^2     &    3.94                     &   3.95
\\ \hline 
\multicolumn{3}{|c|}{
 {\rm  Predicted  \hskip 0.2cm Couplings }}
\\ \hline 
\gamma_{\kappa K\pi}   &  -5.02 \: {\rm GeV}^{-1} &   -5.14 \: {\rm GeV}^{-1}
\\ \hline
\gamma_{\sigma \pi\pi} &  7.26 \: {\rm GeV}^{-1}  &  4.33 \: {\rm GeV}^{-1}
\\ \hline
\gamma_{f\pi\pi}       & 1.46 \: {\rm GeV}^{-1}   & -6.56 \: {\rm GeV}^{-1}
\\ \hline
\gamma_{\sigma K K}    & 9.62 \: {\rm GeV}^{-1}   & 13.69  \: {\rm GeV}^{-1}
\\ \hline
\gamma_{f K  K}    & 10.10 \: {\rm GeV}^{-1}   & -5.78  \: {\rm GeV}^{-1}
\\ \hline   
\end{array}$
\end{center}  
\caption{Extracted parameters from a fit to the $\pi K$ data.
$m_\kappa=897$ MeV.}
\label{Fit_897}
\end{table}

\begin{table}
\begin{center}
$\begin{array}{|c|c|c|}
\hline   
{\rm Fitted \hskip .2cm Parameter}  & {\rm \theta_s=-15.61} & {\rm
\theta_s=86.14}
\\ \hline
   G'_\kappa   &   346 \pm 2 \: {\rm MeV}        &   357 \pm 3\: {\rm  MeV}
\\ \hline   
    M_*        &   1389 \pm 4 \: {\rm MeV}       & 1388 \pm 4 \: {\rm  MeV}
\\ \hline
   \gamma_*    & 4.42 \pm 0.09 \: {\rm GeV}^{-1}   & 4.39 \pm 0.09 \: {\rm GeV}^{-1}
\\ \hline
   G'_*        &  275 \pm 10 \: {\rm MeV}        & 272 \pm 10   \: {\rm MeV}
\\ \hline
   A           & 2.87 \pm 0.03  \: {\rm GeV}^{-1} & 2.96 \pm 0.03  \: {\rm GeV}^{-1}
\\ \hline
   B           &-2.34 \pm 0.03  \: {\rm GeV}^{-1} & -2.56 \pm 0.04  \: {\rm GeV}^{-1}
\\ \hline
  \chi^2     &    3.23                     &   3.26
\\ \hline
\multicolumn{3}{|c|}{
 {\rm  Predicted  \hskip 0.2cm Couplings }}
\\ \hline
\gamma_{\kappa K\pi}   &  -5.75 \: {\rm GeV}^{-1}  &   -5.92 \: {\rm GeV}^{-1}
\\ \hline
\gamma_{\sigma \pi\pi} &  8.36 \: {\rm GeV}^{-1}  &  -4.58 \: {\rm GeV}^{-1}
\\ \hline
\gamma_{f\pi\pi}       & 2.53 \: {\rm GeV}^{-1}   & 8.13 \: {\rm GeV}^{-1}
\\ \hline
\gamma_{\sigma K K}    & 10.45 \: {\rm GeV}^{-1}   & -15.62 \: {\rm GeV}^{-1}
\\ \hline
\gamma_{f K  K}    & 12.76 \: {\rm GeV}^{-1}   & 8.30 \: {\rm GeV}^{-1}
\\ \hline
\end{array}$
\end{center}
\caption{Extracted parameters from a fit to the $\pi K$ data.   
$m_\kappa=875$ MeV.}
\label{Fit_875}
\end{table}

\begin{table}
\begin{center}
$\begin{array}{|c|c|c|}
\hline   
{\rm Fitted \hskip .2cm Parameter}  & {\rm \theta_s=-0.84} & {\rm
\theta_s=71.37}
\\ \hline
   G'_\kappa   &   450 \pm 2 \: {\rm MeV}        &   479 \pm 2 \: {\rm MeV}
\\ \hline   
    M_*        &   1387 \pm 4 \: {\rm MeV}       & 1384 \pm 4  \: {\rm MeV}
\\ \hline
   \gamma_*    & 4.40 \pm 0.09 \: {\rm GeV}^{-1}  & 4.36 \pm 0.09 \: {\rm GeV}^{-1}
\\ \hline
   G'_*        &  273 \pm 10 \: {\rm MeV}        & 268 \pm 11  \: {\rm MeV}
\\ \hline
   A           & 4.32 \pm 0.03  \: {\rm GeV}^{-1} & 4.50 \pm 0.04  \: {\rm GeV}^{-1}
\\ \hline
   B           &-3.91 \pm 0.03 \: {\rm GeV}^{-1}  & -4.29 \pm 0.04 \: {\rm GeV}^{-1}
\\ \hline
  \chi^2     &    2.34                     &   2.39
\\ \hline
\multicolumn{3}{|c|}{
 {\rm  Predicted  \hskip 0.2cm Couplings }}
\\ \hline
\gamma_{\kappa K\pi}   &  -8.64  \: {\rm GeV}^{-1}  &   -9.01  \: {\rm GeV}^{-1}
\\ \hline
\gamma_{\sigma \pi\pi} &  11.76  \: {\rm GeV}^{-1}  &  -4.15 \: {\rm  GeV}^{-1}
\\ \hline
\gamma_{f\pi\pi}       & 7.65 \: {\rm GeV}^{-1}   & 14.52 \: {\rm GeV}^{-1}
\\ \hline
\gamma_{\sigma K K}    & 11.41 \: {\rm GeV}^{-1}   & -20.91  \: {\rm GeV}^{-1}
\\ \hline
\gamma_{f K  K}    & 24.12 \: {\rm GeV}^{-1}   & 19.85 \: {\rm GeV}^{-1}
\\ \hline
\end{array}$
\end{center}
\caption{Extracted parameters from a fit to the $\pi K$ data.   
$m_\kappa=800$ MeV.}
\label{Fit_800}
\end{table}

\begin{table}
\label{a0decay}
\begin{tabular}{lcc}
decay widths & $\Gamma_{a_0}^{tot} = 50 \: {\rm MeV}$ & $\Gamma_{a_0}^{tot} =
100 \: {\rm MeV}$ \\
\tableline 
$\Gamma {\left( {a_0^0} \rightarrow {K^0 \bar {K^0}} \right)}$ & 0.924 MeV
& 2.049 MeV\\ 
$\Gamma \left( a_0^0 \rightarrow K^+ K^- \right)$ & 1.371 MeV & 2.455 MeV\\
\end{tabular}
\caption{Predicted $a_0 \rightarrow K \bar K$ decay widths}
\end{table}

\section{Discussion}

We studied the family relationship of a possible scalar nonet composed
of the $f_0(980)$, the $a_0(980)$ and the $\sigma$ and $\kappa$ type states
found in recent treatments of $\pi\pi$ scattering and $\pi K$ scattering.
The investigation was carried out in the effective Lagrangian framework,
starting from the notion of ``ideal mixing''.  First it was observed that
Okubo's original treatment allows two solutions:  one the conventional
(e.g. vector meson) $q\bar q$ type and the other a ``dual'' picture which is
equivalent to Jaffe's $qq\bar q \bar q$ model.

The four masses of our scalar nonet candidates have a similar, but not
identical pattern to the one expected in the dual ideal mixing picture.  In
order to allow for a deviation from ideal mixing, we have added more terms
to the Lagrangian [see (\ref{mixing-mass-Lag})].  The resulting mass, mixing
and scalar-pseudoscalar-pseudoscalar coupling patterns [see
(\ref{interactions})] were discussed in detail.  The outcome of this
analysis is that the dual picture is in fact favored.  More quantitatively,
the appropriate scalar mixing angle in Eq. (\ref{mixing-convention}) comes
out to be about $-17^{\circ} \pm 4^\circ$ compared with $0^{\circ}$ for
dual ideal mixing and $\pm90^{\circ}$ for conventional ideal mixing.  This
corresponds to $m_\kappa$ ranging from $865-900$ MeV.  

The coupling constant results obtained here may be useful for a number of
applications in low energy hadron phenomenology.  These are defined in
Eq. (\ref{trilinear-interactions}) and listed in Appendix C.  Typical
values of $A$ and $B$ may be read from the small magnitude angle solution
in Tables \ref{Fit_897} and \ref{Fit_875}.  We expect to improve
and further check the accuracy of this model by extending the underlying
models of $\pi\pi$ and $\pi K$ scattering to higher energies and to other
channels.  Finally, it may be interesting to compare our results with those
of quark model and lattice gauge theory approaches to QCD.

\acknowledgments 

The work of D.B., A.H.F. and J.S. has been supported in
part by the US DOE under contract DE-FG-02-85ER 40231.  The work of
F.S. has been partially supported by the US DOE under contract
DE-FG-02-92ER-40704.

\appendix

\section{Diagonalization of Hyperfine Hamiltonian}

In this Appendix, we give some explicit details of the derivation of
(\ref{exotic-hf-mass-splitting}) and (\ref{exotic-mass-eigenstate}) which,
while not being explicitly used in our approach, furnish the main reason
for expecting the scalar $qq\bar{q}\bar{q}$ states to be especially
strongly bound.  Our results agree with those of Jaffe who followed a
different method.  

Let us begin by considering only flavor quantum numbers in order to write down the quark content of members of a $qq\bar{q}\bar{q}$ scalar nonet. Taking the quarks to be in the fundamental representation, $\bf {3}$, of $SU(3)_f$ we have the familiar irreducible decomposition of products of quark states:
\begin{equation}
{\bf {3}} \bigotimes {\bf {3}} = {\bf {6}} \bigoplus {\bf {\bar 3}}
\label{3x3}
\end{equation}
\begin{equation}
{\bf {\bar 3}} \bigotimes {\bf {\bar 3}} = {\bf {\bar 6}} \bigoplus {\bf {3}}.
\label{3barx3bar}\end{equation}
So the only possibility for obtaining a $qq\bar{q}\bar{q}$ flavor nonet is
from the combination $\displaystyle{{\bf {\bar 3}} \bigotimes {\bf {3}}}$
of $q^2 \bigotimes {{\bar q}^2}$ states.  Let $ q_i$ be a basis for the
representation space ${\bf {3}}$, where i=1,2 and 3 correspond to up, down
and strange quarks respectively, with conjugate (antiquark) basis ${\bar
q}^i$.  Then we can consider ``dual quark'' bases corresponding to the $qq$
and $\bar q \bar q$ flavor triplet spaces (thus the states are
antisymmetric with respect to exchange of flavor indices), namely $T_m :=
\epsilon_{mjk}{\bar {q}}^{j}{\bar {q}}^{k}$ and $T^m :=
\epsilon^{mjk}q_{j}q_{k}$.  Up to (anti)symmeterization and linear
combinations we have the flavor nonet given in Eq. (\ref{multiquark-nonet}).
Since $T^m$ and $T_m$ contain at most one strange quark each the nonet
states contain at most two strange quarks.  We note also that, in contrast
to the conventional $q \bar q$ scalar nonet, $N_3^3$ is non-strange in this
realization.

In order to complete the description of $qq\bar{q}\bar{q}$ scalar nonets we
consider the spin and color quantum numbers.  Using the facts that (i) the
$qq$ and $\bar q \bar q$ parts of the state are individually totally
antisymmetric and (ii) the overall $qq\bar{q}\bar{q}$ hadron must be a
color singlet, where the quarks transform according to the fundamental
representation of $SU(3)_c$, we obtain just two possibilities which include
scalar flavor nonets (noting that ${\bf {6_c}} \bigotimes {\bf {{\bar
6}_c}} = {\bf 1_c} \bigoplus {\bf {8_c}} \bigoplus {\bf {27_c}}$), namely
\begin{equation}
{ |0^+,{\bf 9} \rangle }_1:={\left[ 0^{+}, {\bf {\bar {3}}_f}, {\bf {\bar {3}}_c}
\right]}_{qq} \bigotimes {\left[ 0^{+}, {\bf {3_f}}, {\bf {3_c}}
\right]}_{\bar {q} \bar {q}}
\label{nonet1}
\end{equation} 
\begin{equation}
{|0^+,{\bf 9} \rangle }_2:={\left[ 1^{+}, {\bf {\bar {3}}_f}, {\bf {6_c}} \right]}_{qq} \bigotimes {\left[ 1^{+}, {\bf {3_f}}, {\bf {\bar {6}}_c} \right]}_{\bar {q} \bar {q}},
\label{nonet2}
\end{equation}
where we have shown the spin-parity, flavor and color representations respectively for $qq$ and $\bar q \bar q$ separately.  

The ``hyperfine'' interaction Hamiltonian needed for our discussion is
given in Eq. (\ref{hyperfine-Ham}).

Given two representations of SU(n) we have the well-known relationship
between the quadratic Casimirs of these representations, say ${{\bf
J}_A}^2$ and ${{\bf J}_B}^2$, and that of their product:
\begin{equation}
{{\bf J}_A}\cdot{{\bf J}_B}= \frac {1}{2} \left[{{\bf J}_{total}}^2 - {{\bf
J}_A}^{2} - {{\bf J}_B}^2 \right].
\label{casimir}
\end{equation}
 
It can be seen, using (\ref{casimir}), that the parts of the hyperfine
Hamiltonian which involve sums over $qq$ or $\bar q \bar q$ pairs are
diagonal with respect to the bases for the scalar nonets chosen in
(\ref{nonet1}) and (\ref{nonet2}).  In order to calculate the expectation
value of the $q\bar q$ terms in (\ref{hyperfine-Ham}) using (\ref{casimir})
we first expand the bases (\ref{nonet1}) and (\ref{nonet2}) in terms of
states where the spin and color of the $q\bar q$ pairs are explicit. 

To find the recoupling coefficients we follow Close \cite{Close}, where
more detail is given.  For the case of spin recoupling we have, assuming
that all of the quarks in the scalar meson are in relative s-wave states,
that in order to couple to total angular momentum $J=0$, either both $q\bar
q$ pairs must be in $j^P=1^-$ or both in $j^P=0^-$ states, which we denote
as vector, ($V$), and pseudoscalar, ($P$) respectively.  Thus we can expand
the spin part of the state in the following manner:
\begin{equation}
{|J_{total}=0 \rangle }_{1 \: {\rm or} \: 2}  = \alpha PP + \beta VV,
\label{spin}
\end{equation}
where $\alpha$ and $\beta$ can be determined in each case by rewriting both
sides (the left-hand-side will be different for the two states
(\ref{nonet1}) and (\ref{nonet2})) in terms of their constituent quarks and
antiquarks using the usual Clebsch-Gordon identities for $SU(2)$.

Similarly for the color states we note that, since ${\bf 3} \bigotimes {\bf
\bar 3} = {\bf 8} \bigoplus {\bf 1}$, only combinations of the form
\begin{equation}
{\alpha^\prime}|{\bf 8_c} \rangle _{q\bar q} \bigotimes |{\bf 8_c} \rangle
_{q\bar q} + {\beta^\prime}|{\bf 1_c} \rangle _{q\bar q} \bigotimes |{\bf
1_c} \rangle _{q\bar q}
\end{equation}
include color singlets and therefore the color parts of (\ref{nonet1}) and
(\ref{nonet2}) can be written in terms of this basis.  For brevity we
simply present the results of our recoupling coefficient expansions in
Table \ref{recoupling-table}.

\begin{table}[h]
\begin{tabular}{lcc} ${\rm nonet}$ & ${\rm spins\: of} q\bar q {\rm \: pairs}$ &
 ${\rm color \:products \: of} q\bar q {\rm \: pairs}$ \\ \tableline ${|0^+,{\bf
 9} \rangle }_1$ & $\frac{1}{2} PP + \frac {\sqrt 3}{2} VV$ & $\frac {1}{\sqrt 3} {\bf
 1}_c \bigotimes {\bf 1}_c - {\sqrt{\frac{2}{3}}} {\bf 8}_c \bigotimes {\bf
 8}_c$ \\ ${|0^+,{\bf 9} \rangle }_2$ & $\frac{{\sqrt 3}}{2} PP - \frac {1}{2}
 VV$ & $\sqrt {\frac{2}{3}} {\bf 1}_c \bigotimes {\bf 1}_c +
 {\sqrt{\frac{1}{3}}} {\bf 8}_c \bigotimes {\bf 8}_c$ \\
\end{tabular}
\caption{Spin and color recouplings for flavour nonets}
\label{recoupling-table}
\end{table}

\begin{table}[h]
\begin{center}
$\begin{array}{|c|l|} \hline \hline
{\rm {Representation}} & \: {\bf F}^2 \: \\ \hline
{\bf 3} {\: {\rm or} \:} {\bf \bar 3} & \frac {4}{3} \\ 
{\bf 8}& 3 \\ 
{\bf 1} & 0 \\ 
{\bf 6} & \frac {10}{3} \\ \hline \hline
\end{array}$
\end{center}
\caption{SU(3) Quadratic Casimirs}
\label{SU(3)}
\end{table}

In order to give an idea of the next step let us look at one of the
off-diagonal elements of $\langle H_{hf} \rangle$, where $H_{hf}$ is as in
(\ref{hyperfine-Ham}), with respect to the basis given in (\ref{nonet1})
and (\ref{nonet2}).  Labelling the quarks/antiquarks $q_1 q_2 {\bar q}_3
{\bar q}_4$ we have that the only non-vanishing off-diagonal pieces in
$\langle H_{hf} \rangle$ are the sums over $(13)$, $(14)$, $(23)$ and $(24)$.  For
example, applying (\ref{casimir}) yields
\begin{equation}
{\bf S_1 \cdot S_3 F_1 \cdot F_3}|0^+, {\bf9_f}\rangle_1 = \frac {1}{2} \left[ -
{\small {\frac{6}{4} \cdot \frac{1}{2}}}PP +  {\small {\frac{1}{2} \cdot \frac
 {\sqrt 3}{2}}} VV \right] \left[ {\small {- \frac{8}{3} \cdot \frac{1}{\sqrt 3}}} {\bf
 1_c} \bigotimes {\bf 1_c} - {\small {\frac{1}{3} \cdot \sqrt {\frac
 {2}{3}}}} {\bf 8_c} \bigotimes {\bf 8_c} \right],
\end{equation}
where for the color operators we have used the $SU(3)$ Casimirs given in
Table \ref{SU(3)}.  Finally we take the inner product with the expansion of
${|0^+,{\bf 9} \rangle}_2$ in Table \ref{recoupling-table} which gives that 
\begin{equation}
\langle {\bf S_1 \cdot S_3 F_1 \cdot F_3} \rangle_{21} = \frac{1}{4} \sqrt {\frac
{3}{2}}.
\end{equation}
There are, as noted above, four such combinations, all of which contribute
equally.  An analogous calculation can be performed for the diagonal matrix
elements giving finally:

\begin{equation}
\langle H_{hf} \rangle _{ab} = 
- \Delta \left[ \begin{array}{c c}
1 & \sqrt {\frac {3}{2}} \\
\sqrt {\frac {3}{2}} & \frac {11}{6} \end{array} \right], 
\end{equation}   
where a and b run over the indices 1 and 2 labelling the flavor nonets ${|0^+,{\bf
9}\rangle }_1$ and ${|0^+,{\bf 9}\rangle }_2$.  Thus the eigenstates of the hyperfine
interaction correspond to mixtures of these nonets, corresponding to
energies $E_{1} = - 2.71 \Delta$ and $E_{2} = -0.12 \Delta$, which are in
agreement with \cite{Jaffe}.  The corresponding eigenstates are:
\begin{eqnarray}
{|0^+,{\bf 9}\rangle}& =& 0.585 {|0^+,{\bf 9}\rangle}_1 + 0.811{|0^+,{\bf 9}\rangle}_2
\nonumber \\
{|0^+,{\bf 9}^* \rangle} &=& 0.811 {|0^+,{\bf 9}\rangle}_1 - 0.585{|0^+,{\bf 9}\rangle}_2.
\end{eqnarray}

\section{Chiral Covariant Form}

Here we present the terms of the total Lagrangian involving the scalar
nonet $N_a^b(x)$ in chiral invariant or (for the mass terms which break
the chiral symmetry) in chiral covariant form. We follow the general
method of non-linear realization described in \cite{Callan} but our notation
is as in Appendix B of \cite{BFSS}. The object $\xi = {\rm exp}
(i\phi/F_\pi)$ discussed there transforms as 
\begin{equation}
\xi \rightarrow U_L \xi K^\dagger = K \xi U_R^\dagger
\end{equation}
under chiral transformation.  Our nonet field is considered to transform
as
if it were made of ``constituent'' quarks, namely
\begin{equation}
N \rightarrow K N K^\dagger.
\end{equation}
With the convenient objects
\begin{equation}
p_\mu={i\over 2}\left(\xi \partial_\mu \xi^\dagger 
-\xi^\dagger\partial_\mu \xi \right)\hskip .5cm , \hskip .5cm
v_\mu={i\over 2}\left(\xi \partial_\mu \xi^\dagger +
\xi^\dagger\partial_\mu \xi \right)
\end{equation}
we write the additional Lagrangian terms involving $N$:
\begin{eqnarray}
{\cal L} &=& - {1\over 2}{\rm Tr}\left( {\cal
D}_\mu N{\cal D}_\mu N \right) -a {\rm Tr} (N N) -{b\over 2}{\rm Tr} \left[
N N \left( \xi^\dagger {\cal M} \xi^\dagger + \xi {{\cal M}}^{\dagger} \xi \right)
\right]- c {\rm Tr} (N) {\rm Tr} (N) \nonumber \\ 
&-& {d \over 2} {\rm Tr} (N) {\rm Tr} \left[ N \left( \xi^\dagger {\cal M} \xi
^\dagger + \xi {{\cal M}^{\dagger}} \xi \right) \right] + F_\pi^2\left[
A\epsilon^{abc}\epsilon_{def}N_a^d (p_\mu)_b^e(p_\mu)_c^f + B{\rm
Tr}(N){\rm Tr}(p_\mu p_\mu) \right. \nonumber \\ 
&+& \left.  C {\rm Tr} ( N p_\mu) {\rm
Tr}(p_\mu) + D {\rm Tr} (N) {\rm Tr} (p_\mu) {\rm Tr} (p_\mu) \right]
\label{Lagrangian}
\end{eqnarray}
where ${\cal D} = \partial_\mu - i v_\mu$ and $ {\cal M}= {\cal M}^\dagger$
is the spurion matrix defined after (\ref{mass-Lag}).  The entire
Eq.(\ref{Lagrangian}) is formally invariant if we allow ${\cal M}$ to
transform as $ {\cal M} \rightarrow U_L {\cal M} U_R^\dagger$.  This
Lagrangian reproduces (\ref{mixing-mass-Lag}) and
(\ref{interactions}) but also contain interactions with extra pions.  These
extra interactions do not change anything in this paper or in the
tree-level formulas for $\phi\phi$ scattering in
\cite{Harada-Sannino-Schechter} and \cite{BFSS}.

\section{Coupling Constants}
Here we find the scalar-pseudoscalar-pseudoscalar coupling constants
defined in (\ref{trilinear-interactions}) in terms of the parameters $A, B,
C, D$ [see
(\ref{interactions})], the scalar mixing angle [see (\ref{mixing-convention})] and the pseudoscalar
mixing angle, $\theta_p$.  The latter is defined according to:
\begin{equation}
\left( 
\begin{array}{c} 
        \eta\\ 
        \eta' 
\end{array} 
\right) =
\left( 
\begin{array}{c c} 
{\rm cos} \theta_p  & -{\rm sin} \theta_p \\
{\rm sin} \theta_p  &  {\rm cos} \theta_p 
\end{array} 
\right)
\left( 
\begin{array}{c} 
 (\phi^1_1+\phi^2_2)/ \sqrt{2} \\ \phi^3_3 
\end{array} 
\right),
\label{eta-etap}
\end{equation}
where $\eta$ and $\eta'$ are the fields which diagonalize the pseudoscalar
analog of (\ref{mass-matrix}).  The usual convention employs a different
basis; in this convention the angle is $\theta_u$ and
\begin{equation}
\left(
\begin{array}{c}
        \eta\\
        \eta'
\end{array}
\right) =
\left(
\begin{array}{c c}
{\rm cos} \theta_u  & -{\rm sin} \theta_u \\
{\rm sin} \theta_u  &  {\rm cos} \theta_u
\end{array} 
\right)
\left( 
\begin{array}{c} 
(\phi^1_1+\phi^2_2-2\phi^3_3)  / \sqrt{6} \\ 
(\phi^1_1+\phi^2_2 + \phi^3_3) / \sqrt{6} 
\end{array} 
\right).
\label{eta-etap-conventional}
\end{equation}  
The relation between the two angles is 
\begin{equation}
\theta_p=\theta_u + 54.74^o \approx 37^o
\end{equation}
in which case (see for example \cite{Mirelli}) $\theta_u \approx - 18^o$
was taken. More recent analyses (\cite{Schechter-Subbaraman-Weigel} and
\cite{Herrera}) have modified this treatment somewhat by considering
derivative mixing terms as well as non-derivative ones.

Note that the basis for (\ref{eta-etap}) was chosen so that ${\bar q} q$ 
is the more natural picture for the pseudoscalar nonet, in contrast to 
(\ref{mixing-convention}) for the scalars.  Because the mixing angles can take on any
values, this in no way biases the analysis one way or the other.

The $\gamma$'s are predicted in the present model as
\begin{eqnarray}
\gamma_{\kappa  K \pi} & = & \gamma_{a_0 K K} = -2 A
\label{a_0-coupling}  \label{kappa-K-pi} \\
\gamma_{\sigma \pi \pi} &=& 
2 B {\rm sin}\theta_s  - \sqrt{2} (B-A) {\rm cos }\theta_s 
\\
\gamma_{\sigma K {K}} &=& 
2 (2 B - A) {\rm sin}\theta_s -2 \sqrt{2} B {\rm cos }\theta_s  
\\
\gamma_{f_0 \pi \pi} &=& 
\sqrt{2} (A - B) {\rm sin }\theta_s - 2 B {\rm cos}\theta_s
\\
\gamma_{f_0 K {K}} &=& 
2 (A - 2 B) {\rm cos}\theta_s -2 \sqrt{2} B {\rm sin }\theta_s  
\label{f-K-barK}\\
\gamma_{\kappa {K} \eta} &=&  
C {\rm sin} \theta_p - \sqrt{2} (C-A) {\rm cos }\theta_p  
\\
\gamma_{\kappa {K} \eta '} &=&  
\sqrt{2} (A-C) {\rm sin }\theta_p -  
C {\rm cos} \theta_p  
\\
\gamma_{a_0 \pi \eta} &=&  
(C- 2A) {\rm sin }\theta_p -  
\sqrt{2} C {\rm cos} \theta_p  
\\
\gamma_{a_0 \pi \eta'} &=&  
(2A - C) {\rm cos }\theta_p -  
\sqrt{2} C {\rm sin} \theta_p  
\\
\gamma_{\sigma\eta\eta} &=&
\left[ \sqrt{2}(B+D) -{1\over 2}(C + 2A + 4D) {\rm sin}2\theta_p
+ \sqrt{2}(C+D) {\rm cos}^2\theta_p \right] {\rm sin}\theta_s
\nonumber \\ &&
- 
\left[ (B+D) -{1\over \sqrt{2} }(C + 2 D) {\rm sin}2\theta_p
+ (A+D) {\rm cos}^2\theta_p  + C {\rm sin}^2\theta_p\right] 
{\rm cos}\theta_s
\\
\gamma_{\sigma\eta'\eta'} &=&
\left[ \sqrt{2}(B+D) + {1\over 2}(C + 2A + 4D) {\rm sin}2\theta_p
+ \sqrt{2}(C+D) {\rm sin}^2\theta_p \right] {\rm sin}\theta_s
\nonumber \\ &&
- 
\left[ (B+D) + {1\over \sqrt{2} }(C + 2 D) {\rm sin}2\theta_p
+ (A+D) {\rm sin}^2\theta_p  + C {\rm cos}^2\theta_p\right] 
{\rm cos}\theta_s
\\
\gamma_{\sigma\eta\eta'} &=&
\left[ \sqrt{2}(C+D){\rm sin}2\theta_p + (C + 2A + 4D) {\rm cos}2\theta_p
\right] {\rm sin}\theta_s
\nonumber \\ &&
- 
\left[ \sqrt{2} (C+2 D) {\rm cos}2 \theta_p + (A - C + D)
{\rm sin}2\theta_p \right] 
{\rm cos}\theta_s
\\
\gamma_{f_0\eta\eta} &=&
\left[- \sqrt{2}(B+D)+ {1\over 2}(C + 2A + 4D) {\rm sin}2\theta_p
- \sqrt{2}(C+D) {\rm cos}^2\theta_p \right] {\rm cos}\theta_s
\nonumber \\ &&
- 
\left[ (B+D) -{1\over \sqrt{2} }(C + 2 D) {\rm sin}2\theta_p
+ (A+D) {\rm cos}^2\theta_p  + C {\rm sin}^2\theta_p\right] 
{\rm sin}\theta_s
\\
\gamma_{f_0\eta'\eta'} &=&
-\left[ \sqrt{2}(B+D)+ {1\over 2}(C + 2A + 4D) {\rm sin}2\theta_p
+ \sqrt{2}(C+D) {\rm sin}^2\theta_p \right] {\rm cos}\theta_s
\nonumber \\ &&
- 
\left[ (B+D) +{1\over \sqrt{2} }(C + 2 D) {\rm sin}2\theta_p
+ (A+D) {\rm sin}^2\theta_p  + C {\rm cos}^2\theta_p\right] 
{\rm sin}\theta_s
\\
\gamma_{f_0\eta\eta'} &=&
-\left[ \sqrt{2}(C+D){\rm sin}2\theta_p + (C + 2A + 4D) {\rm
cos}2\theta_p
\right] {\rm cos}\theta_s
\nonumber \\ &&
- 
\left[ \sqrt{2} (C+2 D) {\rm cos}2 \theta_p + (A - C + D)
{\rm sin}2\theta_p \right] 
{\rm sin}\theta_s
\end{eqnarray}

\end{document}